\documentclass[usenatbib,usegraphicx,useAMS]{mn2e}

\title[Optical Emission from SNR G156.2+5.7]
	{Discovery of Extensive Optical Emission Associated with the X-ray Bright, 
         Radio Faint Galactic SNR G156.2+5.7}

\author[Gerardy \& Fesen]{Christopher L.~Gerardy$^{1}$ \& Robert 
A.~Fesen$^2$\\ 
$^1$Astrophysics Group, Imperial College London, Blackett Laboratory, 
Prince Consort Road, London SW7 2AZ, UK\\
$^2$Department of Physics \& Astronomy, Dartmouth College, 6127 Wilder
Laboratory, Hanover, NH 03755-3528, USA}

\begin{document}
\date{13 Aug 2006}

\pagerange{\pageref{firstpage}--\pageref{lastpage}} \pubyear{2006}

\maketitle

\label{firstpage}

\begin{abstract}

We present wide-field H$\alpha$ images of the Galactic supernova remnant
G156.2+5.7 which reveal the presence of considerable faint H$\alpha$ line
emission coincident with the remnant's X-ray emission. We also present
low-resolution optical spectra for a few representative emission regions.  The
outermost H$\alpha$ emission consists largely of long and thin (unresolved),
smoothly curved filaments of Balmer-dominated emission presumably associated
with the remnant's forward shock front. Patches of brighter H$\alpha$ emission
along the western, south-central, and northeastern regions appear to be
radiative shocked ISM filaments like those commonly seen in supernova remnants,
with relatively strong [O~I] $\lambda\lambda$6300,6364 and [S~II]
$\lambda\lambda$6716,6731 line emissions. Measured [S~II]
$\lambda\lambda$6716,6731/H$\alpha$ ratios range from 0.84 to 1.57.

Comparison of the observed H$\alpha$ emission with the {\it ROSAT} PSPC X-ray
image of G156.2+5.7 shows that the thin Balmer-dominated filaments lie along
the outermost edge of the remnant's detected X-ray emission. Brighter radiative
emission features are not coincident with the remnant's brightest X-ray or
radio regions. Areas of sharply weaker X-ray flux seen in the {\it ROSAT} image
of G156.2+5.7 appear spatially coincident with dense interstellar clouds
visible on optical and {\it IRAS} 60 and 100 ~$\mu$m emission images, as well
as maps of increased optical extinction. This suggests significant X-ray
absorption in these regions due to foreground interstellar dust, especially
along the western and southern limbs. The close projected proximity and
alignment of the remnant's brighter, radiative filaments with several of these
interstellar clouds and dust lanes hint at a possible physically interaction
between the G156.2+5.7 remnant and these interstellar clouds and may indicate a
smaller distance to the remnant than previously estimated.   

 \end{abstract}


\section{Introduction}

A majority of the $\sim$230 currently confirmed Galactic supernova remnants
(SNRs) listed in the latest SNR catalogue \citep{Green04} were first identified
through radio surveys. However, with the advent of sensitive X-ray satellites,
about half a dozen remnants have now been discovered via their X-ray emission
(cf. \citealt{Busser96} and references therein).

The first Galactic SNR discovered through its X-ray emission was 
reported by \citet*{Pfeff91}. They found a previously unknown Galactic SNR
based on the {\it ROSAT} detection of a relatively bright and nearly circular
108' diameter emission structure centred on $\alpha$(J2000) = 4$^{\rm h}$
59$^{\rm m}$ $7^{s}$, $\delta$(J2000) = 51$^{\rm o}$ $46'$ $35''$. With an
estimated 0.1 -- 2.4~keV X-ray flux  of $1.9 \times 10^{-10}$ erg cm$^{-2}$
s$^{-1}$, this new remnant ranked among the ten brightest X-ray Galactic SNRs
known at that time.  \citet{Pfeff91} concluded that the remnant was located in
a region of very low interstellar density (0.01 atoms cm$^{-3}$), which might help 
account, in part, for the remnant's lack of strong radio emission.

The new remnant, designated G156.2+5.7 (also RXJ04591+5147), was
subsequently detected and confirmed in the radio by \citet*{Reich92}.  In
contrast to the object's bright X-ray flux, its 1 GHz radio surface brightness
of $5.8 \times 10^{-23}$ W m$^{-2}$ Hz$^{-1}$ sr$^{-1}$ ranks it as perhaps the
faintest Galactic SNR currently known \citep{Green04}, being a factor of two in
1 GHz surface brightness below than of the exceptionally radio faint but
optically [O~III] bright remnant, G65.3+5.7 \citep{Gull77,Reich79}. Follow-up
searches for a central radio pulsar or central compact object have proven
unsuccessful \citep{Lorimer98,Kaplan06}.

\begin{figure*} 
\includegraphics[width=170mm]{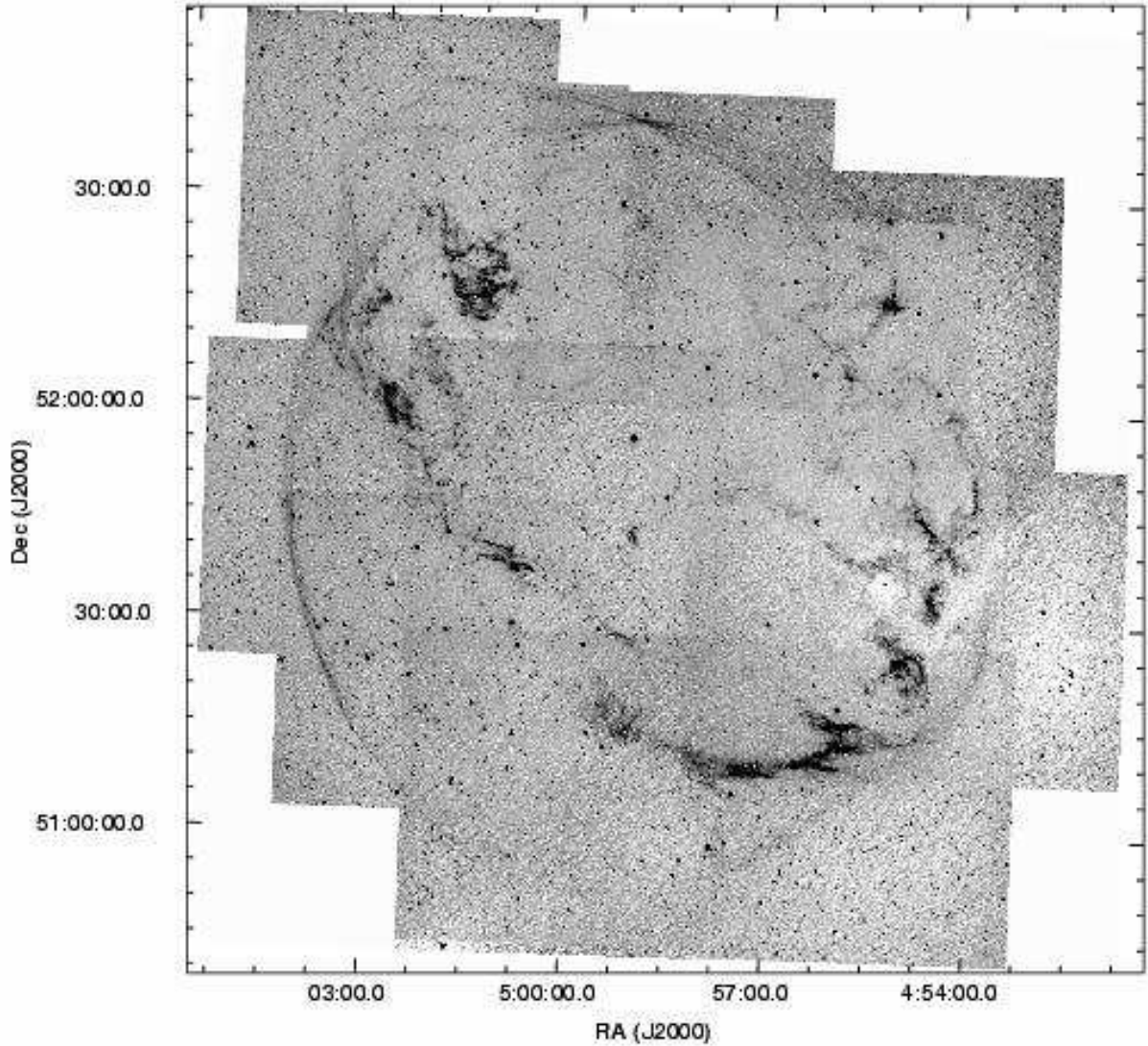} 
\caption{
H$\alpha$\ mosaic image of the G156.2+5.7 supernova remnant.  Two distinct
emission morphologies are seen: a faint rim of thin filamentary emission which
runs most of the way around the remnant; and patches of much brighter emission
with a clumpy, complicated structure. Note the white dust absorption patch
along the remnant's western limb. To reduce the appearance of background noise
the data have been smoothed slightly using a Gaussian filter.  
\label{ha}} 
\end{figure*} 

Since shortly after its discovery, G156.2+5.7 has been studied only in the
X-rays. Using the {\it ASCA} satellite, \citet{Yama99,Yama00} found both hard
and soft X-ray components, an estimated Sedov age of $\sim 1.5 \times 10^{4}$
yr, and an ambient gas density of $\sim$0.2 cm$^{-3}$. More recently,
\citet{Pann04} used both {\it ASCA} and {\it RXTE} to study the the
northwestern rim of G156.2+5.7 and investigate the presence of a non-thermal
X-ray component and cosmic-ray acceleration.

\citet{Pfeff91} reported finding no associated optical emission on the Palomar
Sky Survey plates and there are no discernible emission features in this region
visible on the relatively deep, narrow passband filter images of
\citet*{Parker79} in their emission-line survey of the Milky Way.  However,
here we present wide field H$\alpha$ images of the remnant which reveal a
remarkably extensive and complex optical emission structure covering a large
fraction of the remnant's X-ray shell.

\section{Observations}
\subsection{Narrow-band H$\alpha$ Imaging} 
Wide-field imaging of G156.2+5.7 (hereafter referred to as G156) was performed
using the McDonald Observatory 0.76-m telescope and the Prime-Focus Corrector
(PFC) camera \citep{Claver92}.  This system uses a 2048$\times$2048 Loral
Fairchild front-side illuminated CCD which produces a plate scale of
$1\farcs35$ per pixel and an unvignetted field of view of
$46\farcm1\times46\farcm1$ with little spatial distortion ($<0\farcs5$,
far better than the typical seeing).

Narrow passband imaging was carried out using a pair of matched interference
filters (FWHM 30 \AA) centred at 656.8 nm (= H$\alpha$) and 651.0 nm (=
adjacent red continuum).  These filters were optimised for an f/7 setup, which
caused some concern given the faster (f/3) optics of the 0.76-m/PFC setup.
However, tests on-sky suggest that the faster optics did not dramatically
affect the performance.  In particular we detect no significant change in
H$\alpha$ sensitivity between the centre and the edge of the detector (other
than the variation due to the flat-field).  The H$\alpha$ setup suffers from a
strong internal reflection which is not present in the continuum images, and
results in a characteristic halo around bright point sources.  

Dome flats for both filters were obtained nightly and, whenever clear, twilight
flats were also obtained in morning and evening twilight.  Since the S/N of the
dome flats was significantly better, they were the primary images used for
flat-field correction.  However, the twilight flats were used to perform
illumination corrections on the nightly dome flats.  The illumination image was
created via a two-stage process.  First, all the twilight flats were combined,
and each individual twilight flat was divided by the resulting average twilight
image.  Any twilight frames which showed a significant gradient or other
structure relative to the average were then removed and a second image was
created using the reduced set of twilight frames.  Each of the nightly dome
flats was then divided by this refined twilight image, and the illumination
correction for the dome flat created via a 2-D, low-order polynomial fit.  

Observations of G156 were carried out over three nights, 12, 18, \& 19 January
2004, in conditions ranging from photometric to moderate cirrus.  Since the
remnant is much larger than the PFC field of view, it was covered in 13
partially overlapping telescope pointings.  For each pointing a pair of 1000~s
exposures was obtained in both the H$\alpha$ and 6510 continuum filters.  As
the PFC does not have a working guider, the images were unguided, and the
tracking rate of the telescope manually adjusted using an empirical mapping of
the tracking rate for different declinations and hour angles.  Inevitably, this
open-loop tracking and the long exposures required for narrow-band imaging
resulted in some images with relatively poor point-spread functions (PSFs).  At
the scale of the remnant, this degradation of the image quality is not a
serious problem.  However, the PSF of the paired H$\alpha$ and continuum images
are not always a perfect match if the tracking drift was larger in one of the two
images.  Even in well tracked image pairs, the internal reflections in the
H$\alpha$ filter leave a residual halo around bright stars in the continuum
subtracted images.

After bias and flat-field correction, each pair of H$\alpha$ and continuum
images were registered via cross correlation routines in IRAF, and trimmed to
the overlapping region.  The H$\alpha$ images obtained in non-photometric data
were scaled to match the flux levels in the photometric frames by matching the
flux in aperture photometry of stars in the overlapping regions between the
different pointings.  For a couple of fields, there were no adjacent pointings
taken in photometric conditions, and the flux was set matching re-scaled
H$\alpha$\ flux in an adjacent image. The relative flux across the entire
remnant is probably accurate to $\sim 20$\%.  

After setting the flux levels for each H$\alpha$\ image, the paired continuum
image was scaled to match using aperture photometry of a large number of field
stars.  The continuum image was then subtracted from the H$\alpha$\ image at
each pointing, and the resulting continuum subtracted image was combined into a
single mosaic.  The continuum subtraction process often left a small residual
level in the background, which was removed before the images were mosaiced
together.  This reduced the `patchwork quilt' appearance of the final mosaic,
although variations in signal-to-noise and small flat-fielding errors mean that
the combination of these multiple pointings is not seamless (see Fig.\
\ref{ha}).  The reduced 6510~\AA\ images were also similarly combined to create
a mosaic continuum image.

\subsection{Optical Spectroscopy}
\begin{figure}
\includegraphics[width=84mm]{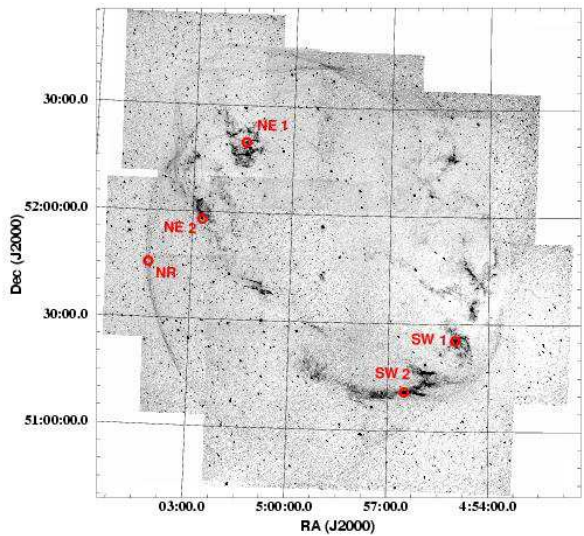}
\caption{H$\alpha$ image of the G156 SNR with the locations of the observed 
optical spectra shown.
\label{finder}}
\end{figure}

Low-dispersion optical spectra of a few of the remnant's brighter emission
knots were obtained on three nights between 10 -- 12 September 2004 using the
MDM 2.4-m telescope at Kitt Peak Arizona and the Modular Spectrograph with a
600 lines mm$^{-1}$ 6000 \AA \ blaze grating and a $1\farcs 5 \times 4'$ slit.
Spectra were taken at four locations in the remnant's detected optical
emission, two in the NE and SW (see Fig.~\ref{finder}) with the slit orientated
north-south across relatively bright features of the detected optical emission. 

Total exposure times ranged from 300 to 900~s yielding spectra with an
effective coverage of 6000 -- 8000 \AA\ and a spectral resolution of $\simeq$
2~\AA.  Standard IRAF software was used for the data reduction with wavelength
calibration using Hg, Ne, and Xe lamps and flux calibration via
\citet{Massey90} standard stars.  Although all three nights were photometric,
seeing varied between $1\farcs 0 - 1\farcs 5$. As a result, slit light losses
were considerable at times due to the variable seeing conditions and guiding
errors. Consequently, our measured absolute flux values are only accurate to
$\pm$ 25\%.

\begin{figure*}
\includegraphics[width=150mm]{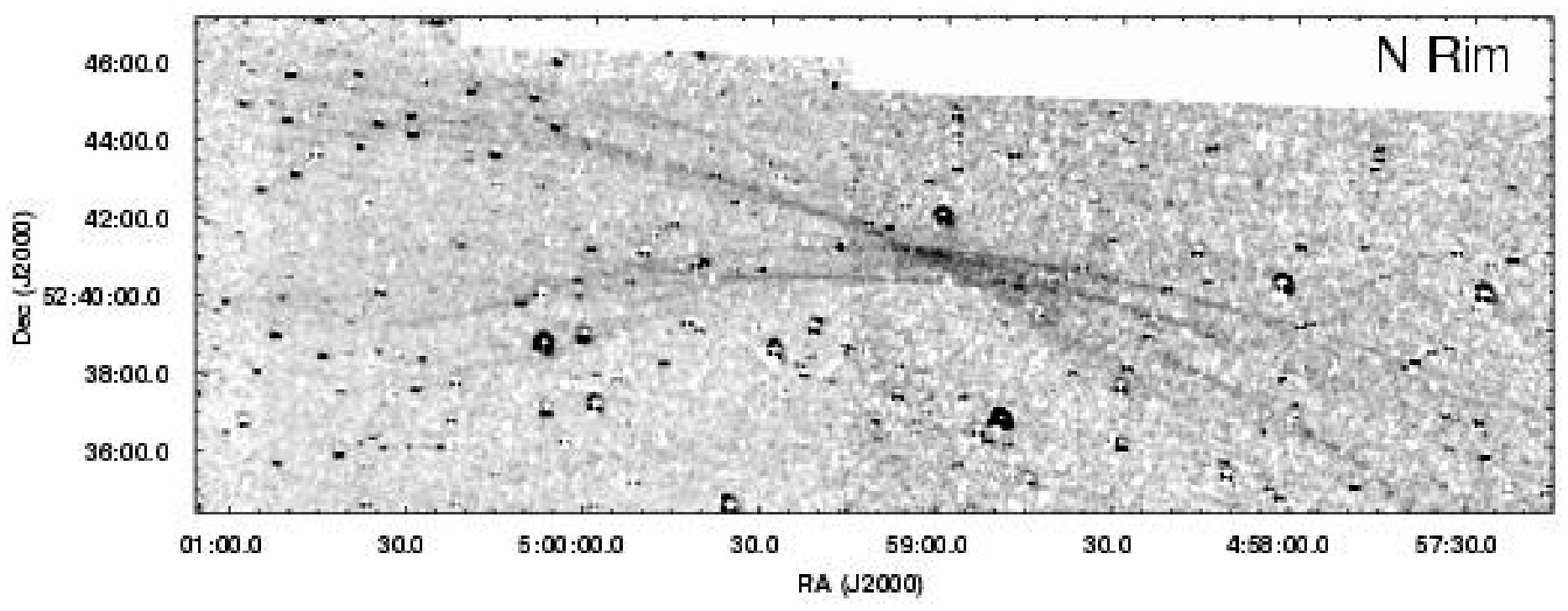}\\
\includegraphics[width=54mm, clip=True, trim=0bp 0bp 0bp 10bp]{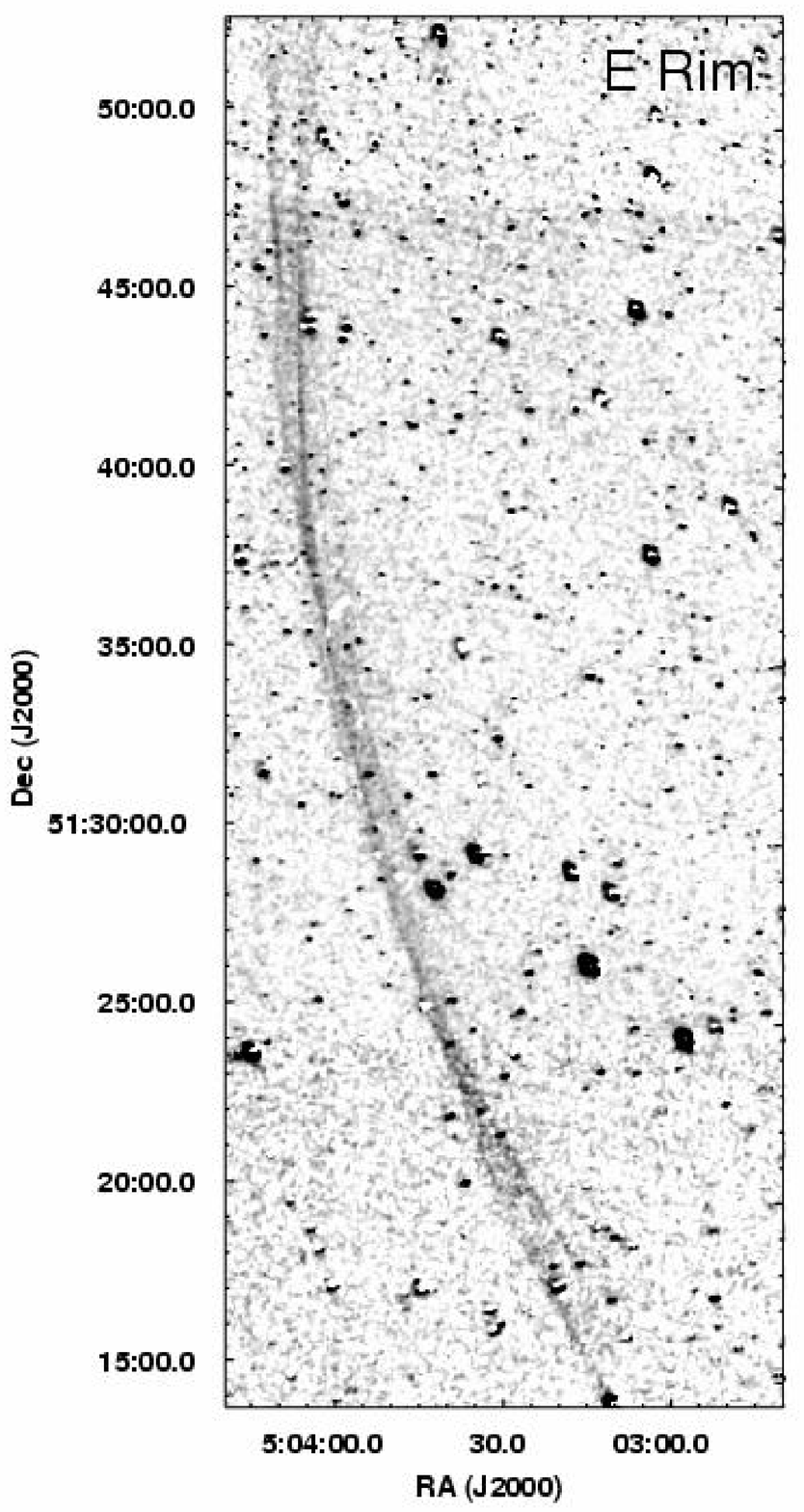}
\includegraphics[width=112mm, clip=True, trim=0bp 40bp 0bp 0bp]{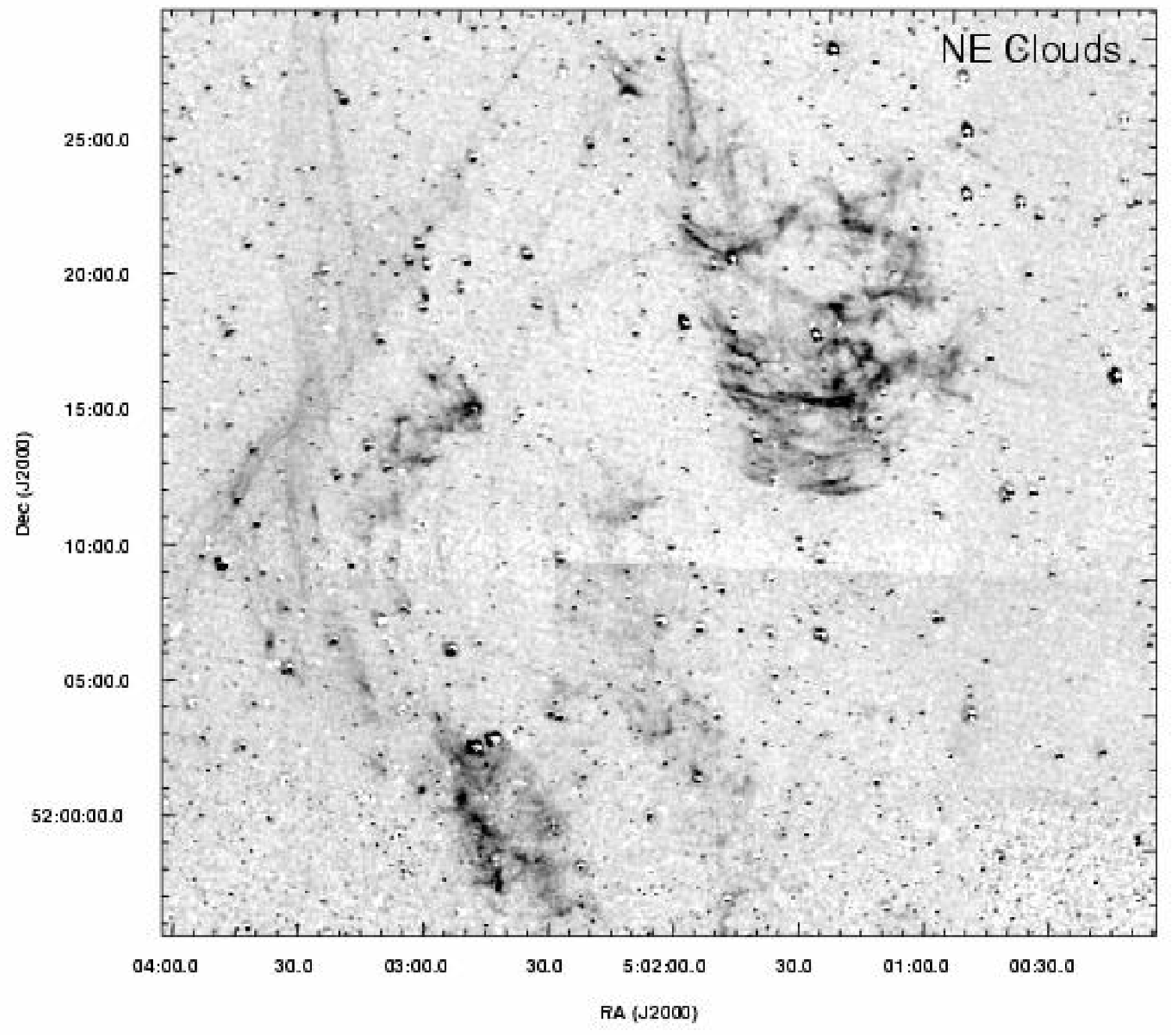}\\
\caption{
Enlargements of the H$\alpha$ mosaic image showing thin, overlapping
filamentary arcs near the middle of the northern rim (top), and eastern rim
(left) of G156.  These filaments are likely non-radiative shocks producing
Balmer-dominated emission.  A mixture of thin non-radiative features and
brighter emission filaments can been seen near the northeast corner of the
remnant (right).  To reduce the appearance of background noise the data have
been smoothed using a Gaussian filter.
\label{blowups}}
\end{figure*}

A separate 2000~s spectrum was taken of the suspected Balmer-dominated,
non-radiative filament along the remnant's eastern limb.  These data were
obtained on 9 October 2005 under non-photometric conditions using the MK~III
spectrograph on the MDM 2.4-m telescope.  A $1\farcs 7 \times 4'$ east-west
aligned slit together with a 300 lines mm$^{-1}$ 5000 \AA \ blaze grism was
used to yield a spectrum coverage of 5500 -- 8000 \AA, calibrated as before
with Hg, Ne, and Xe lamps and \citet{Massey90} standard stars.

\section{Results}
\subsection{H$\alpha$ Mosaic Image}
Our continuum subtracted H$\alpha$\ mosaic image is shown in Figure~\ref{ha}.
The mosaic shows an extensive and complex H$\alpha$\ emission structure.  While
numerous features are readily apparent in this image, they are actually quite
faint, with an estimated surface brightness of $\leq 2 \times 10^{-16}$ erg
cm$^{-2}$ arcsec$^{-2}$ s$^{-1}$ for most of the brighter regions. The optical
emission seen here is invisible on the Palomar Sky Survey, in the relatively
deep narrow passband filter images of \citet{Parker79}, and is not listed in
the Atlas of Galactic Nebulae \citep{NV85}.  

Two distinct morphologies are seen; faint, thin (often unresolved) filamentary arcs
some tens of arc minutes long, and much brighter more diffuse filamentary
features like those seen in many optical Galactic SNRs.  The thin filaments are
primarily found near the edges of the detected optical emission structure, and
are brightest along the eastern limb ($\alpha = 5^{\rm h}03^{\rm m} - 5^{\rm
h}04^{\rm m}$, $\delta =  51\degr15' - 51\degr50'$) and the northern limb 
($\alpha = 4^{\rm h}58^{\rm m} - 5^{\rm h}01^{\rm m}$, $\delta = 52\degr40'$)
of the G156 SNR.  

Two regions exhibiting thin and partially overlapping filamentary arcs are
shown in greater detail in the upper and lower left panels of
Figure~\ref{blowups}.  The morphology of such filamentary emission is
reminiscent of the non-radiative, Balmer-dominated shock emission along the
outer edges of supernova remnants (e.g., the Cygnus Loop, Tycho's SNR, and SN
1006). Indeed, a spectrum of these filaments along the east rim of the detected
emission structure (at the point labelled ``NR'' in Figure~\ref{finder}) showed
only H$\alpha$ line emission within the 6000 -- 8000 \AA \ wavelength range.
Such a spectrum is consistent with it being a Balmer filament associated with
non-radiative shock emission. Although brightest to the north and east, these
filaments can be seen to form a nearly unbroken shell of emission surrounding
the brighter, more diffuse emission regions.  These thin filaments fade out in
the northwest near $\alpha = 04^{\rm h}55^{\rm m}$, $\delta = 52\degr15'$ and
in the south at $\alpha =$5$^{\rm h}00^{\rm m}$, $\delta = $ 50$\degr55'$.
                                                                                                                    
The lower right panel of Figure~\ref{blowups}\ shows examples of both the faint
filamentary emission and much brighter, more diffuse and clumpy filament type
emission.   Clumpy filaments are the brightest H$\alpha$ features detected in
the G156 region and form a patchy arc which loops from the NE down to the SW
region of the remnant region (see Fig.\ 1).  In the NE, this arc is dominated
by a large complex of filaments at $\alpha = 5^{\rm h}01^{\rm m} - 
5^{\rm h}02^{\rm m}$, $\delta = 52\degr10' - 52\degr25'$, which
exhibits considerable internal structure.  We will refer to this feature as the
``NE filament complex''. In the southwest portion of the H$\alpha$ mosaic
image, this clumpy type of emission forms a bright ridge which is broken by an
apparent foreground dust-lane near $\alpha = 4^{\rm h}54^{\rm m} - 4^{\rm h}55^{\rm m}$,
$\delta = 51\degr25' - 51\degr45'$.  Small patches of clumpy
emission are also seen in the NW near the break in the outer thin filamentary
emission, and near the centre of the image.  

\begin{figure}
\includegraphics[width=85mm]{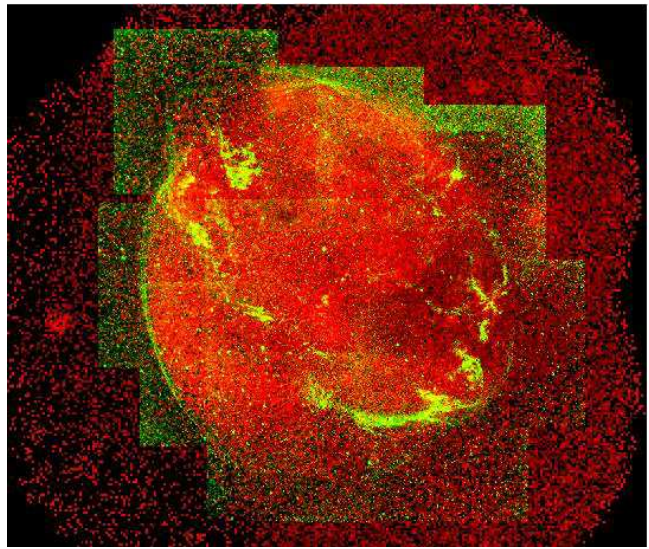}
\caption{
The X-ray image (red) of G156 from the {\it ROSAT} PSPC 2$\degr$\ Survey 
overlaid with our 
continuum subtracted H$\alpha$\ image (green). The X-ray and H$\alpha$ images
are displayed using a square-root and log intensity scales, respectively.  The
non-radiative H$\alpha$\ filaments lie just beyond the outer edge of the X-ray
emission.  A weak bulge of X-ray emission in the NE is similarly bounded by
faint H$\alpha$ filaments extending outward from the otherwise quite circular
boundary of G156.
\label{xray_ha}}
\end{figure}

\subsection{Comparison of Detected H$\alpha$ Emission with X-ray and Radio
Emission} 
Comparison of the observed H$\alpha$\ emission with a {\it ROSAT} PSPC
$2\degr$\ Survey X-ray image of G156 suggests that virtually all of the
detected optical emission is associated with the G156 remnant.
Figure~\ref{xray_ha}\ shows the {\it ROSAT} PSPC X-ray image (0.1 -- 2.4~keV)
in red with our H$\alpha$\ emission mosaic superimposed in green.  

This figure shows that the detected gentle arcs of thin filamentary emission
lie along the outermost edge of the remnant's observed X-ray emission,
confirming that these are indeed non-radiative emission from the outer blast
wave of the supernova remnant. Along the NE limb, the non-radiative filaments
bulge outward a bit from their otherwise relatively circular arc. This bulge in
the optical corresponds well to faint extended X-ray emission seen in the {\it
ROSAT} image.  The rim of H$\alpha$\ emission fades out close to the region in
the southeast where the X-ray emission is also faint.  In contrast, the clumpy
optical emission is less clearly associated with emission structures in the
X-ray image, but appears to lie somewhat near the boundaries of darker (X-ray
faint) regions in the {\it ROSAT} image.
                                                                                                                    
\begin{figure*}
\includegraphics[width=80mm]{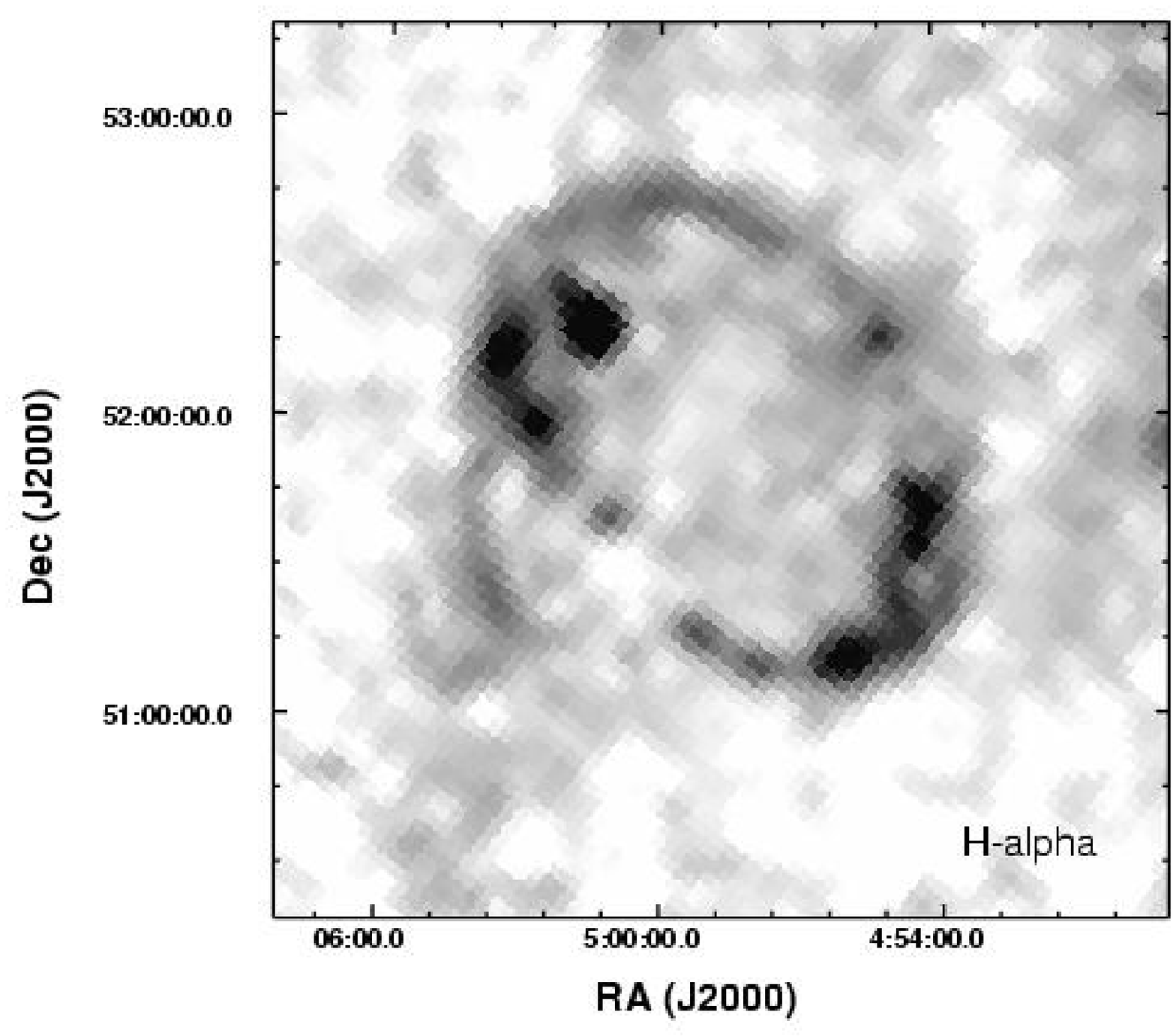}
\includegraphics[width=80mm]{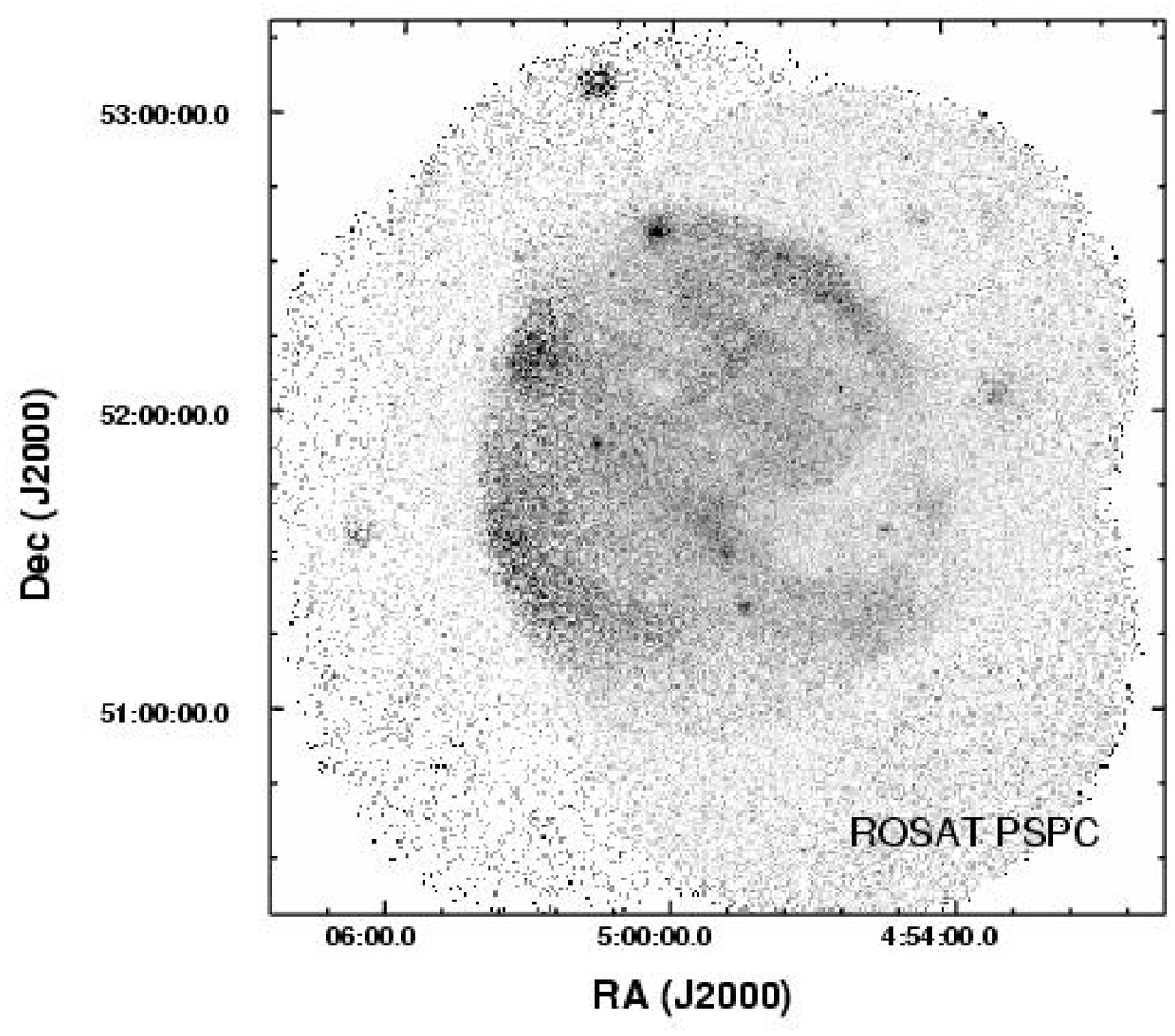}\\
\includegraphics[width=80mm]{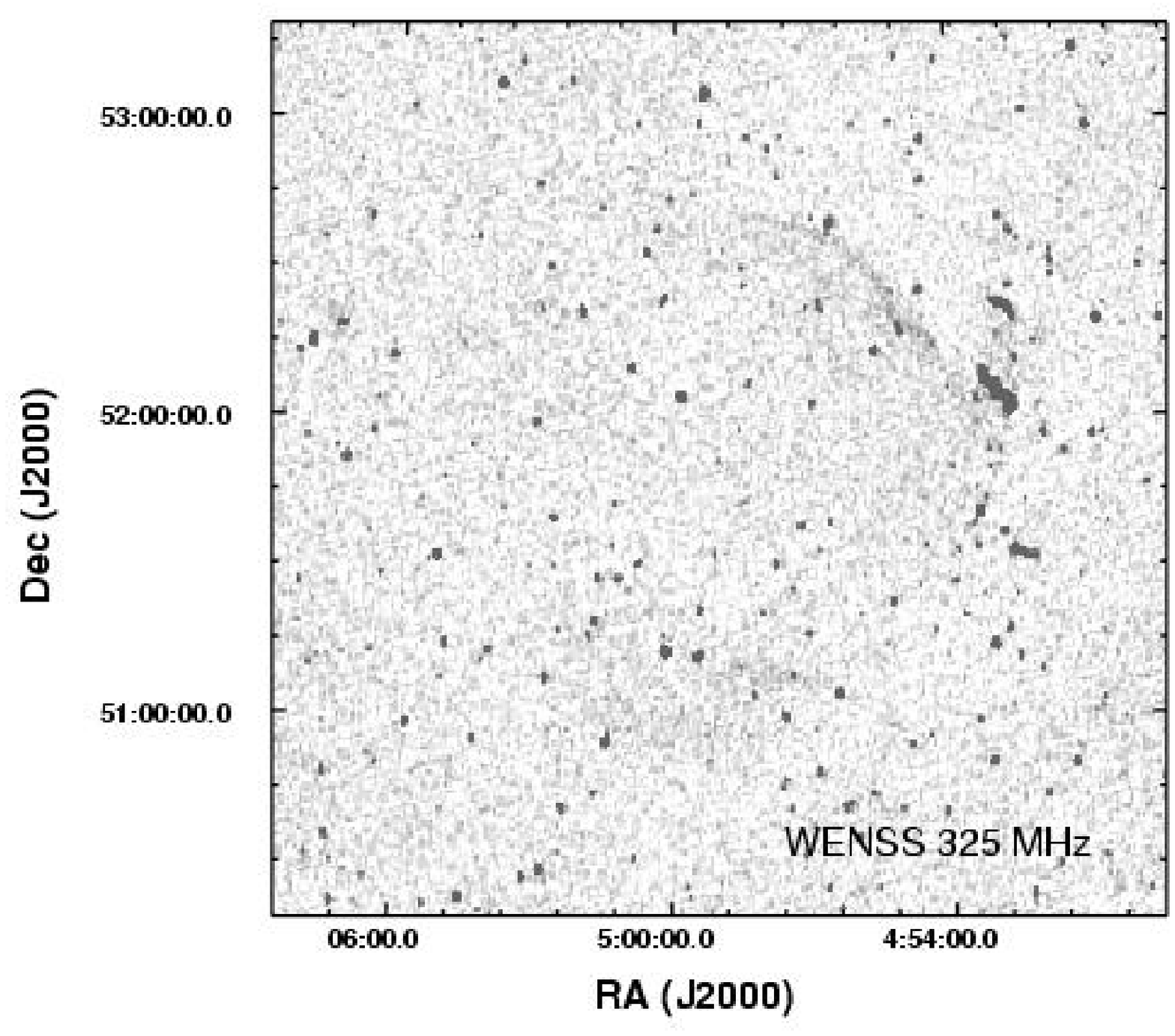}
\includegraphics[width=80mm]{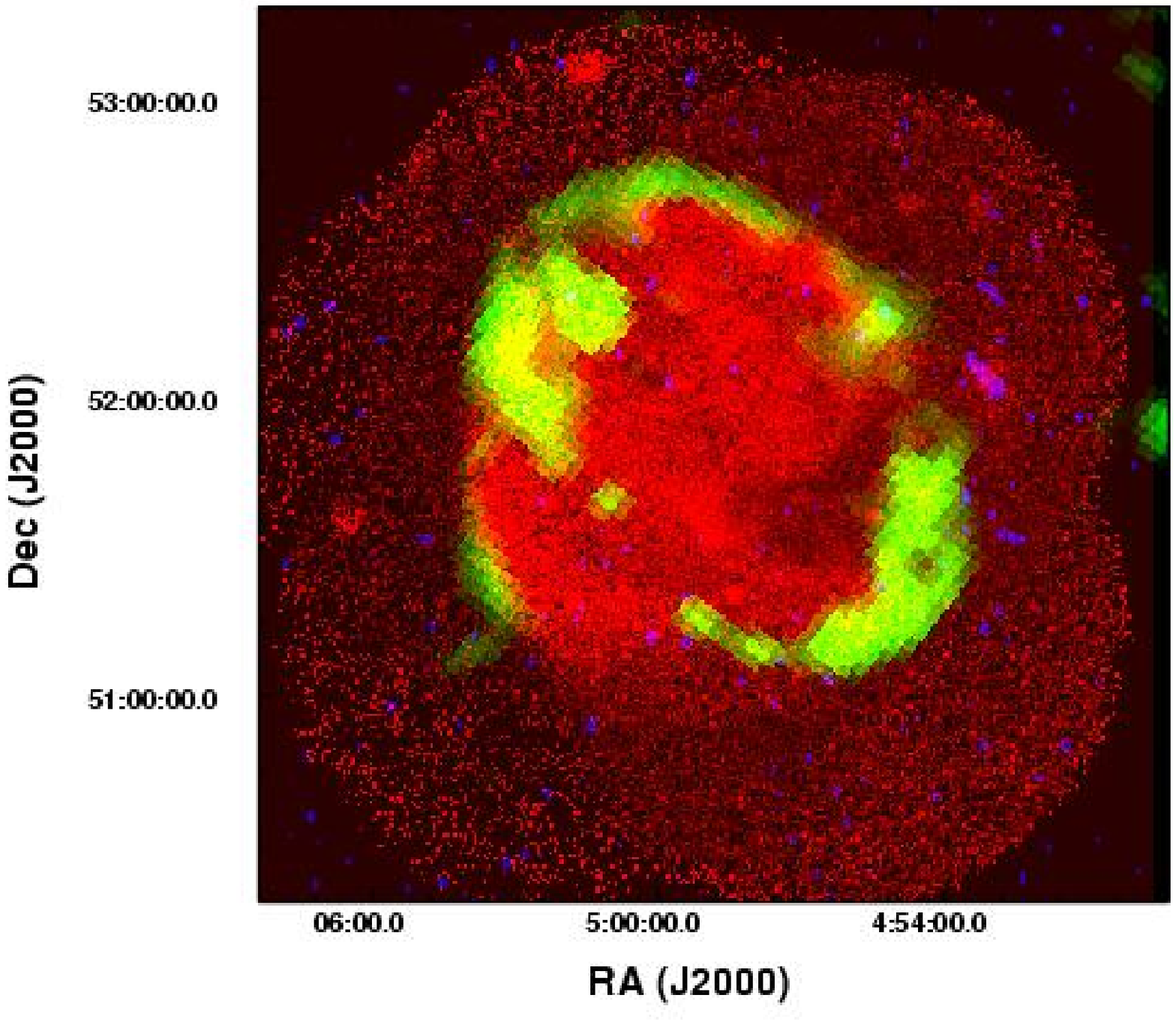}
\caption{
Upper left: Low spatial resolution H$\alpha$\ image of G156 from the Virginia
Tech Spectral Line Survey \citep[VTSS;][]{Fink03}.  Upper Right: X-ray (0.1 --
2.4 keV) image of G156 from the {\it ROSAT} PSPC 2$\degr$ Survey.  Lower Left:
325~MHz image of G156 from the Westerbork Northern Sky Survey
\citep[WENSS;][]{Rengelink97}.  Lower Right: Colour-composite of these three
images with X-rays in red, H$\alpha$\ in green, and 325~MHz in blue.
\label{skyview}}
\end{figure*}

As a follow-up to our detection of H$\alpha$ in this region, we conducted an
on-line search of survey data of the G156 region with the \textit{Skyview
Virtual Observatory}\footnote{http://skyview.gsfc.nasa.gov} \citep{McGlynn96}.
Surprisingly, the H$\alpha$ emission we found associated with G156 was, in
fact, well detected (but unreported) by the Virginia Tech Spectral Line Survey
(VTSS) \citep[][Fig.~\ref{skyview}, upper left]{Fink03}, albeit at much lower
spatial resolution.  Comparison of our H$\alpha$\ mosaic with the VTSS image
shows good agreement between the two data sets after allowing for the much
lower spatial resolution of the VTSS image.  

Small portions of the G156 remnant can also be weakly seen in the 325~MHz
Westerbork Northern Sky Survey \citep[WENSS][Fig.~\ref{skyview}, lower
left]{Rengelink97}, showing a similar radio structure to that seen at 1448 and
2695 MHz \citep{Reich92}.  The remnant's radio emission appears nearly
anti-correlated with the optical Balmer filaments, as it shows only two faint
filaments to the south where the Balmer-dominated rim emission is weakest, 
and to the northwest where non-radiative filaments are poorly
detected (although there is faint diffuse H$\alpha$ emission and a bright clump
of diffuse emission here.) The VTSS H$\alpha$, {\it ROSAT} X-ray, and WENSS
325~MHz images are all shown in Figure~\ref{skyview} -- in the upper left,
upper right and lower left, respectively -- with all three overlaid as a three-colour
image in the lower right.

\subsection{Optical Spectra of G156 Filaments}
\begin{figure*}
\includegraphics[width=65mm,angle=270]{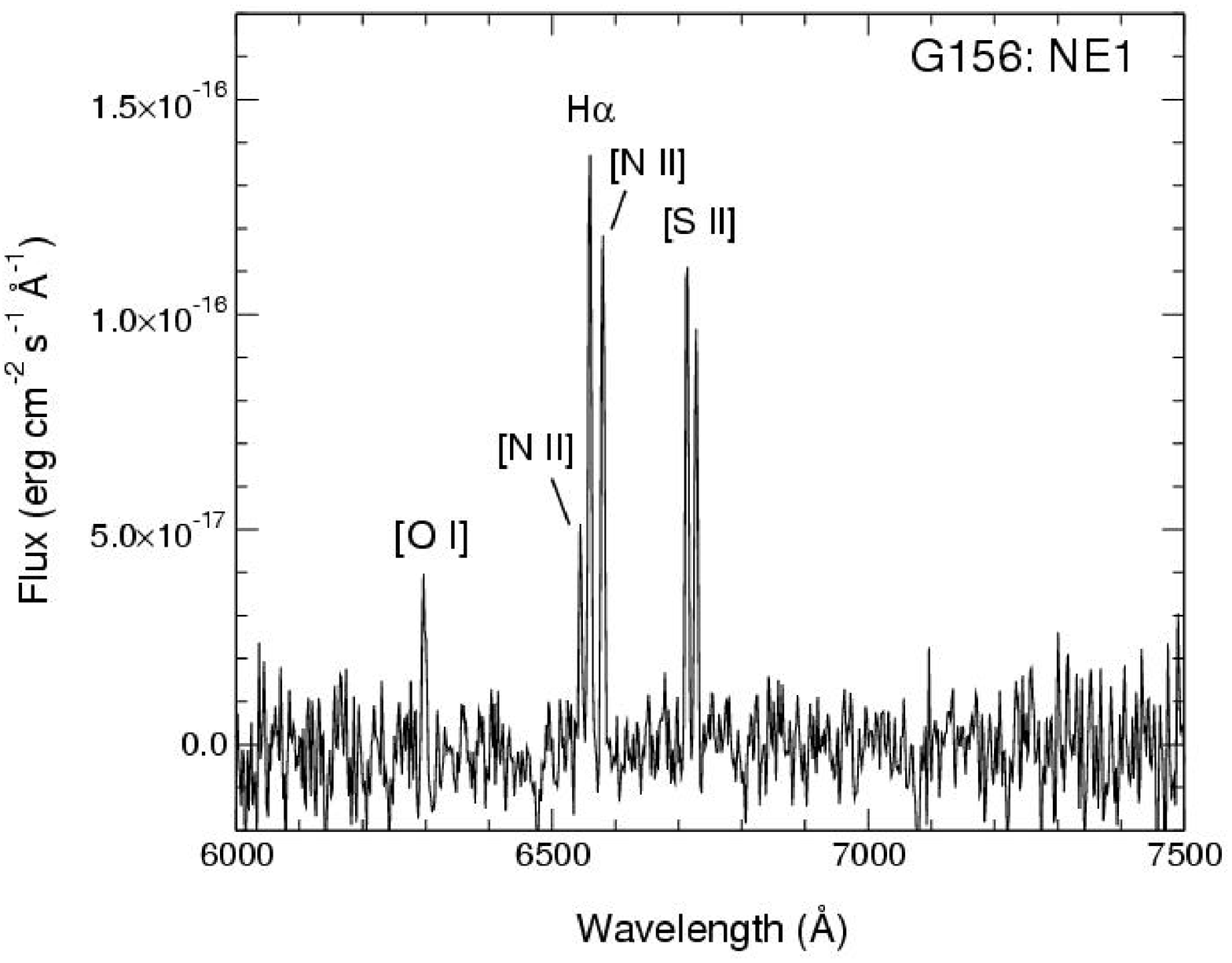}
\includegraphics[width=65mm,angle=270]{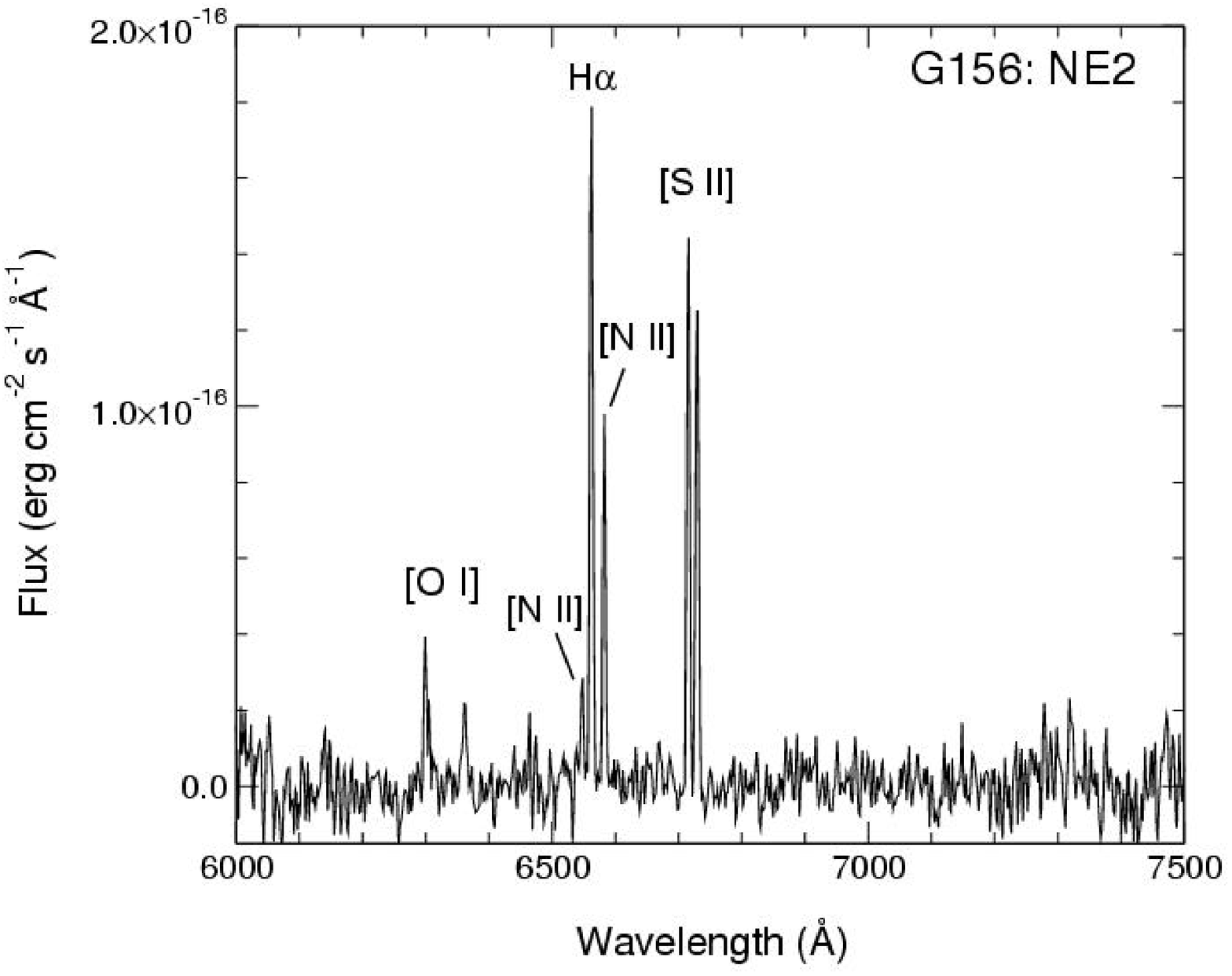}\\
\includegraphics[width=65mm,angle=270]{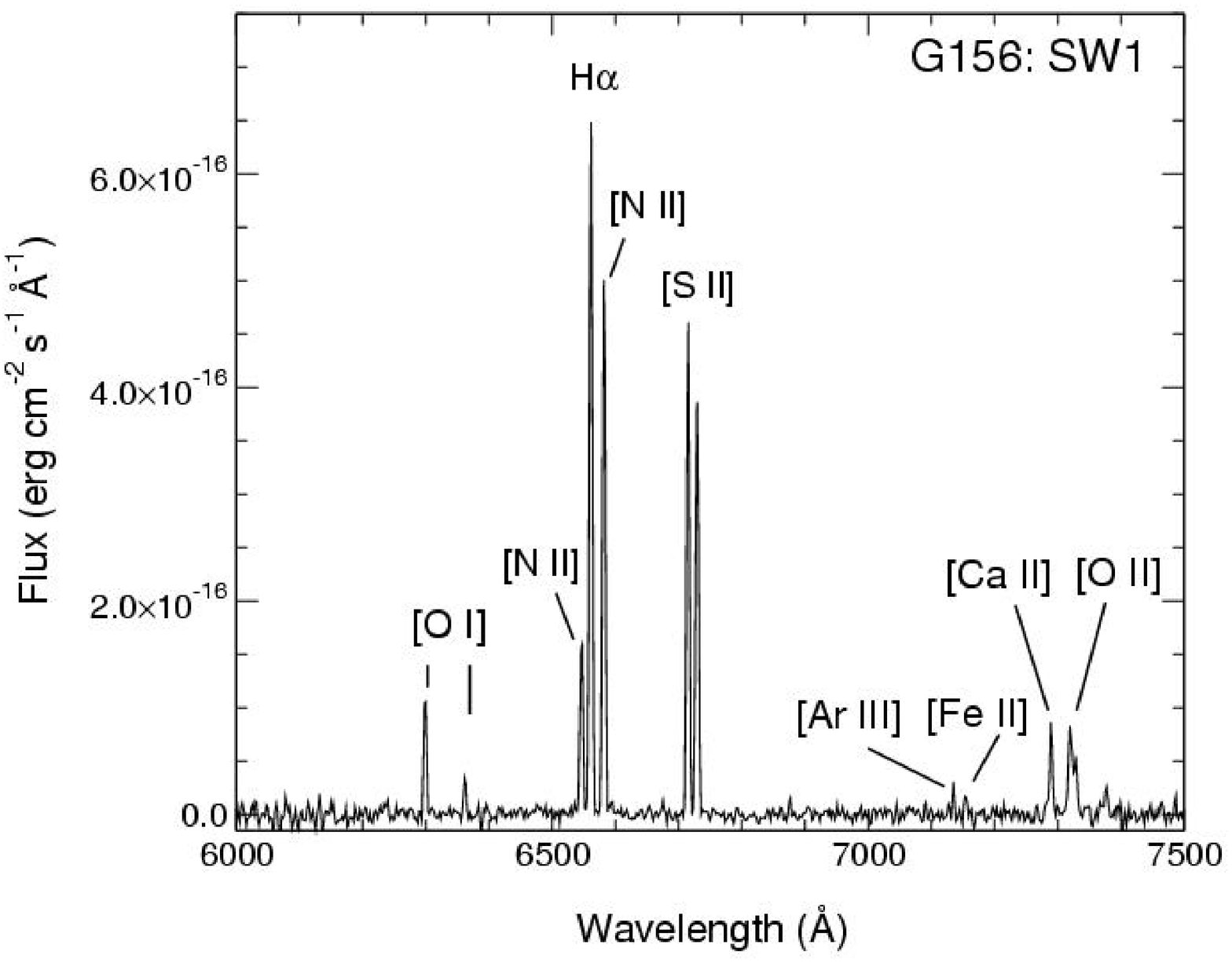}
\includegraphics[width=65mm,angle=270]{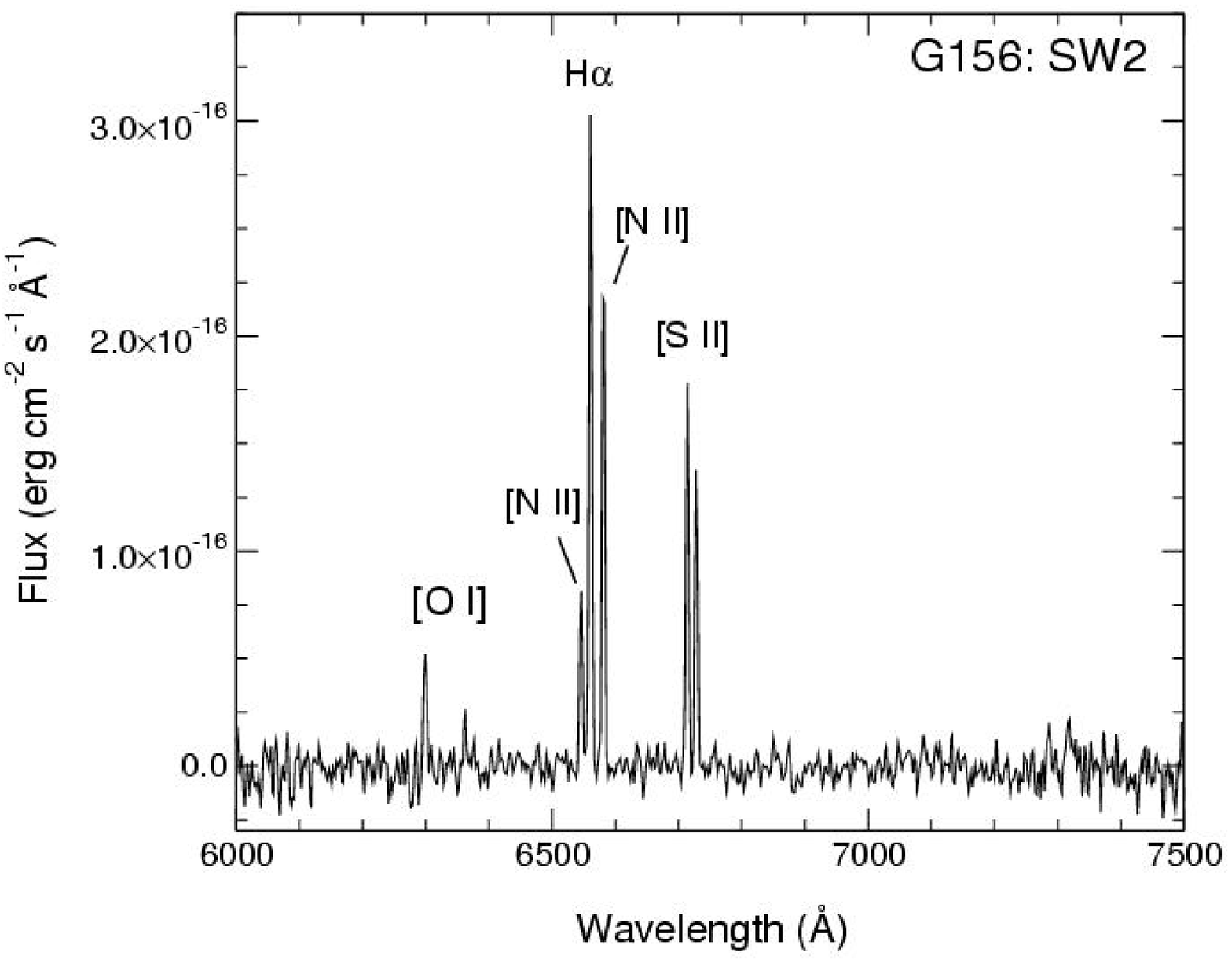}
\caption{Optical spectra of emission from radiative shocks at four locations
in G156.  The locations are labelled in Figure~\ref{finder}.
\label{spectra}}
\end{figure*}

\begin{table}
 \caption{Observed Relative Line Intensities}
\label{symbols}
\begin{tabular}{@{}cccccc}
\hline
 Line   &  ID       &   NE1  &   NE2   &   SW1     &   SW2     \\
\hline
6300    & [O I]     &  ~88   &   ~62   &   ~50     &   ~45     \\
6548    & [N II]    &  108   &   ~40   &   ~80     &   ~85     \\
6563    & H$\alpha$ &  300   &   300   &   300     &   300     \\
6583    & [N II]    &  270   &   145   &   233     &   245     \\
6716    & [S II]    &  225   &   250   &   213     &   135     \\
6731    & [S II]    &  180   &   220   &   180     &   115     \\
7135    & [Ar III]  &  ...   &   ...   &   ~10     &   ...     \\
7155    & [Fe II]   &  ...   &   ...   &   ~~9     &   ...     \\
7291    & [Ca II]   &  ...   &   ...   &   ~36     &   ~20     \\
7319,30 & [O II]    &  ...   &   ...   &   ~54     &   ~30     \\
        &           &        &         &           &           \\
6716/6731    & [S II] ratio     &  1.25  &  1.14   &  1.18     & 1.16      \\
6716,31/6563 & [S II]/H$\alpha$ &  1.35  &  1.57   &  1.31     & 0.84      \\
6300,63/6563 & [O I]/H$\alpha$  &  0.40  &  0.27   &  0.22     & 0.20      \\
        &           &        &         &           &           \\
             & H$\alpha$ Flux$^{a}$   & 8.9& 10 &  36  & 30   \\
             & N$_{e}$$^{a}$    &  175   &  325    &  270      & 300       \\
\hline
\end{tabular}\\
$^{a}$ H$\alpha$ flux is listed in units of 10$^{-16}$ erg cm$^{-2}$ s$^{-1}$.  \\
$^{b}$ Electron densities are given in units of cm$^{-3}$. \\
\end{table}

The optical spectra of four filaments of G156's brighter emission labeled in
Figure~\ref{finder}\ are shown in Figure~\ref{spectra}.  Observed line fluxes
and ratios from these data are listed in Table~\ref{symbols}.  The individual
emission line profiles for H$\alpha$ and [S~II] were unresolved at a spectral
resolution of 2 \AA, with no radial velocities observed above 150 km s$^{-1}$.

The well established criteria for identifying shock emission is a
[S~II]/H$\alpha$\ ratio greater than 0.4 \citep{Raymond79,Fesen85,BL97}.  All
four G156 spectra exhibit much larger [S II]/H$\alpha$\ ratios than this
threshold (0.8 -- 1.6; see Table 1), strongly indicating that the detected
emission filaments are shocked interstellar gas. Likewise, the relatively
strong [O~I] $\lambda\lambda$6300,6364 emission seen in these spectra also
in line with shock emission, since [O~I] line emission is quite weak relative to
H$\alpha$ in H~II regions and photoionized nebulae \citep{Fesen85}.

Overall, the observed spectra are typical of radiative emission in supernova
remnants, even down to the presence of [Ar III] $\lambda$7135, [Fe~II]
$\lambda$7155, and [Ca~II] $\lambda$7291 \ seen in the spectrum of SW1.  In
summary, both the morphology of the brighter emission nebulae seen in the G156
region and the observed spectra of these nebulae suggest that these emissions
constitute radiative shocked material associated with the G156 supernova remnant.

Electron densities, N$_{e}$, for each spectrum were calculated using the
measured $\lambda$6716/$\lambda$6731 [S~II] ratio values and the Space
Telescope Science Data Analysis System task ``nebular.temden'' which is based
on a five-level atom approximation and assuming T = 10$^{4}$ K.  The observed
[S II] 6716/6731~\AA \ line ratios imply moderately high post-shock densities
N$_{e} = 200 - 300$ cm$^{-3}$.  Such densities imply a much higher ambient
density than the estimates from analysis of the X-ray emission $n_0 = 0.01$ --
0.2~cm$^{-3}$ \citep{Pfeff91,Yama99,Yama00}.  Assuming a shock velocity
$\sim$100 km s$^{-1}$, a [S~II] derived electron density of 300 cm$^{-1}$
suggests a preshock density of around 10 cm$^{-1}$ \citep{Fesen80}. The obvious
suggestion is that G156 is interacting with a clumpy interstellar medium
resulting in patches of radiative emission where the blast wave is running into
dense clouds, and showing only a Balmer-dominated rim of emission in much 
lower-density regions.
                                                                                                                    
\begin{figure*}
\includegraphics[width=170mm,clip=True,trim=0bp 120bp 0bp 120bp]{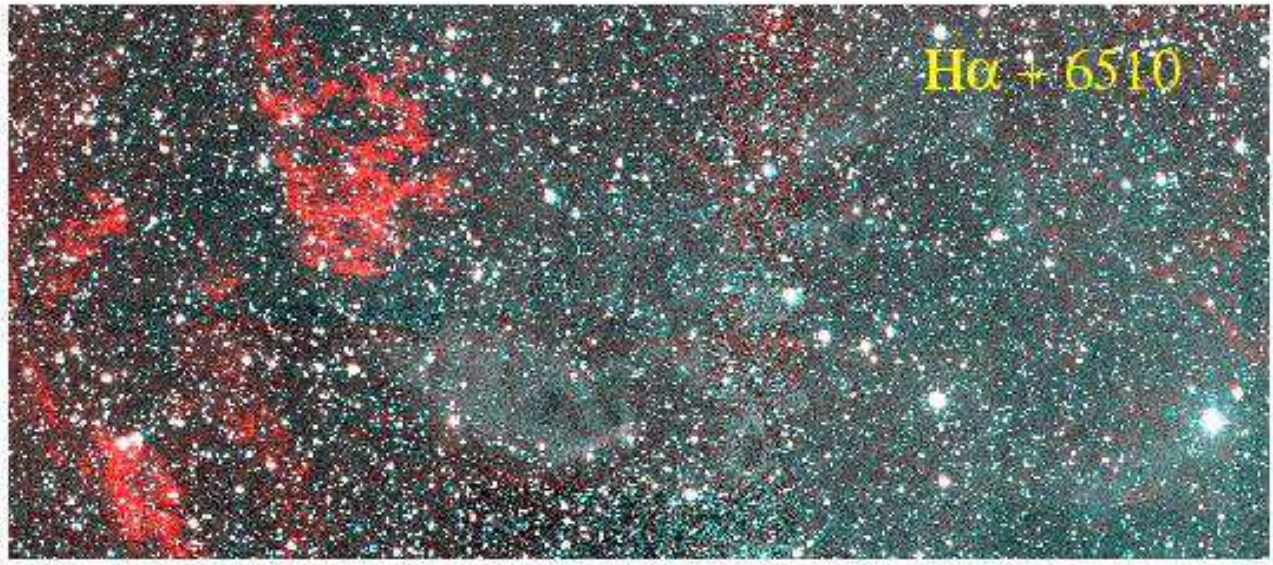}
\caption{
A composite image showing both the continuum 6510~\AA\ image (greyscale) and
the continuum subtracted H$\alpha$ image (red) for the region near the ``NE
filament complex''.  The bright clump of H$\alpha$ emission appears near a
large dark cloud which is seen in the image as both faint continuum emission
and a lack of background stars.  The gap between the southern end of the
H$\alpha$\ emission and the northeastern end of the dark cloud also shows faint
continuum emission.  This apparent spatial coincidence is perhaps suggestive of
a cloud-supernova remnant interaction.
\label{clouds}}
\end{figure*}

\subsection{ISM Clouds in the G156 Region} There appear to be several dense
interstellar clouds along the line of sight to G156. Figure~\ref{clouds}\ is a
composite image showing both the 6510~\AA\ continuum mosaic (greyscale) and the
continuum-subtracted H$\alpha$\ (red) mosaic for a section in the northeastern
portion of the remnant.  The bright H$\alpha$\ emission near the left of the
image was referred to earlier as the ``NE filament complex''.  The lower part
of this emission complex appears about $10' - 20'$ NE of a dark interstellar
cloud which appears in the 6510~\AA\ image as a region lacking in background
stars but with a faint continuum emission, presumably due to ambient radiation
scattered by dust near the surface of the cloud.  Very weak continuum emission
is detected extending from the upper NE end of the cloud which appears to end
at approximately the southern extent of the NE filament complex's H$\alpha$\
emission.  The larger opaque cloud visible just below the centre of this image is a
known Bok globule ($\alpha$[J2000] $ =$4$^{\rm h}59^{\rm m}50.74^{\rm s}$,
$\delta$[J2000] $ = $ 52$\degr04'43.8''$; LDN 1439; CB 26;
\citealt{CB88,CYH91}) and harbors a young stellar object (YSO) with
an associated Herbig-Haro object located some $6.15'$ off to the northwest 
\citep[HHO~494;][]{Stecklum04}. 

The projected spatial coincidence between shocked radiative emission in the NE filament
complex and this dusty interstellar cloud makes it tempting to suggest that part of the 
remnant is actually colliding with the outskirts of this cloud.  The bright
H$\alpha$\ emission to the SW appears similarly spatially coincident with a
large region of high extinction which results in the dark dust lane seen in
Figure~\ref{ha}.  However, there is no clear evidence in the optical images for
significant dust obscuration near some of the other radiative emission features.

\begin{figure}
\includegraphics[width=85mm]{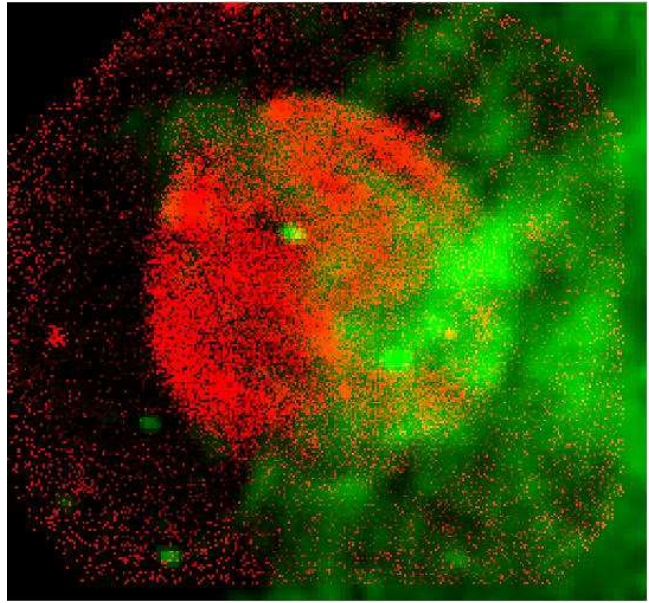}
\caption{
X-ray image (red) of G156 from the {\it ROSAT} PSPC 2$\degr$\ Survey overlaid
with the {\it IRAS} 60 micron image (green) for the same region. Both images
are displayed using a square-root intensity scale.  This shows the excellent
correlation between the warm ISM cloud regions and several major features seen
in the X-ray image due to foreground dust absorption.  The bright {\it IRAS}
point source in the SW quadrant of G156 is coincident with the T-Tauri star
V347 Aur.
\label{xray_iras}}
\end{figure}

\begin{figure}
\includegraphics[width=85mm]{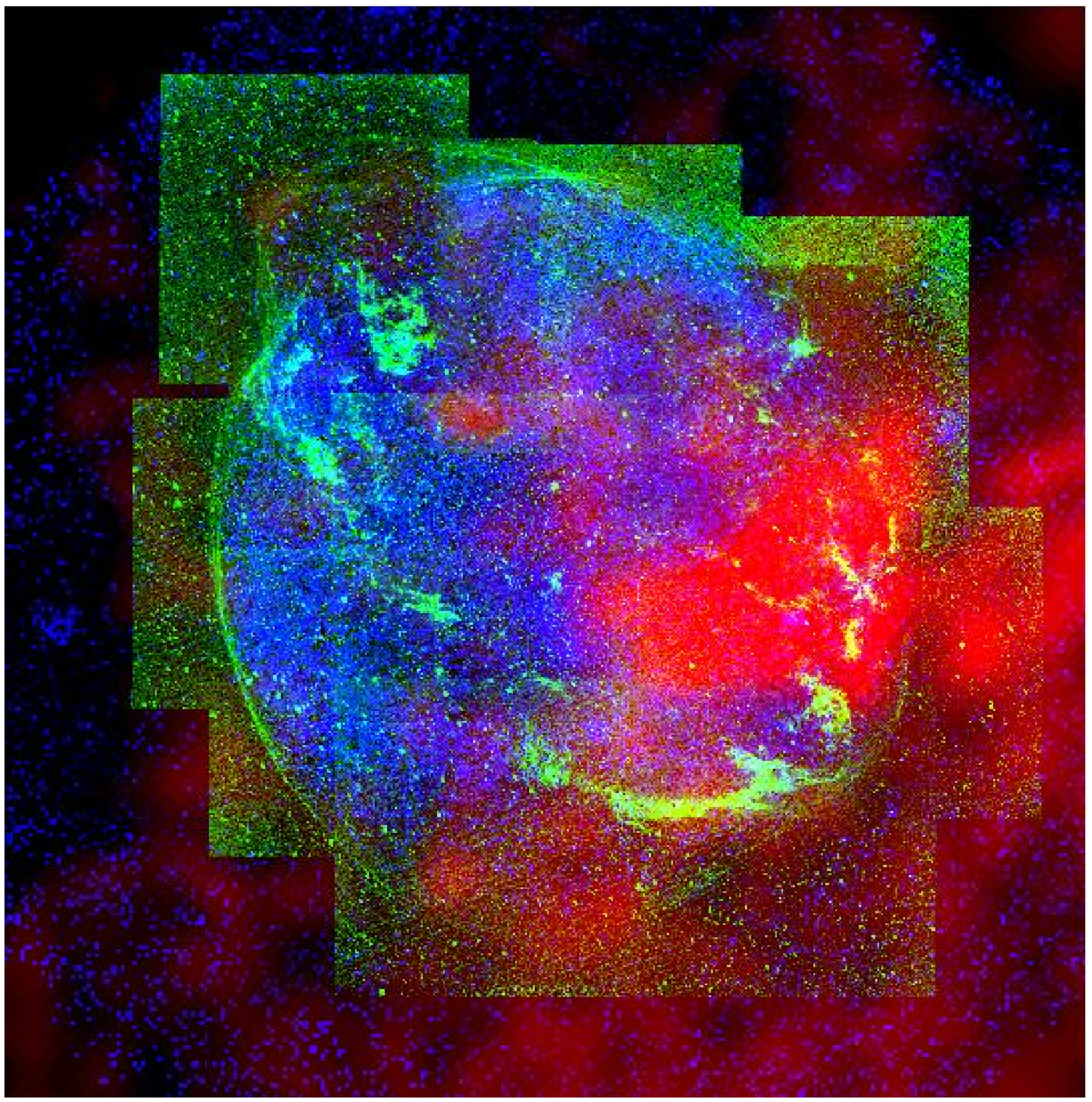}
\caption{
X-ray image (blue) of G156 from the {\it ROSAT} PSPC 2$\degr$\ Survey overlaid
with our H$\alpha$ mosaic (green) and the 6' resolution $A_V$ extinction map
(red) of \citet{Dobashi05}. The Both the X-ray and H$\alpha$ images are
displayed using a square-root intensity scale.  The $A_V$ map is shown on a log
scale and peaks around $A_V=3$.  This image shows the excellent correlation
between the foreground dust clouds and major features seen in the X-ray image.  
\label{xray_Av}}
\end{figure}

However, there is clear evidence that much of the structure in the {\it ROSAT}
image is due to X-ray absorption from these interstellar dust clouds.  The dark
clouds near the NE filament complex and along the southwest rim of G156 are
exactly coincident with dark patches seen in the X-ray image. In
Figures~\ref{xray_ha} and \ref{skyview}, the NE radiative filament complex
appears as a bright green feature near the NE rim of the remnant, with the dark
interstellar cloud to its southwest visible as a virtual `hole' in the X-ray
emission.  Similarly, along the southwest edge of the remnant, a large patch of
significantly decreased X-ray emission can be seen in the {\it ROSAT} X-ray
image coincident with a large dust lane visible in the H$\alpha$\ mosaic image
(Fig.\ 1). This western patch of faint X-ray emission extends well into the
interior of the remnant and is coincident with a region showing several dusty
interstellar clouds.  One of these dusty clouds (dia. = $5'$) is associated with
the young T-Tauri star V347 Aur ($\alpha$[J2000] $ =$4$^{\rm h}56^{\rm
m}56.7^{\rm s}$, $\delta$[J2000] $ = $ 51$\degr30'56''$; RNO 33;
\citealt{Cohen78,Cohen80,OB95} in LDN~1438 \citep{Lynds62} at an estimated
distance of around $250 - 300$ pc.

The coincidence between the dark ISM clouds and the X-ray emission `holes' can
be clearly seen in Figures~\ref{xray_iras}\ and \ref{xray_Av}.  In
Figure~\ref{xray_iras}, we compare the {\it ROSAT} X-ray image (red) with the
{\it IRAS} 60~$\mu$m image (green).  The dark clouds appear in emission in the
{\it IRAS} image and correlate  very well with both the cloud near the NE
filament complex, and the large triangular shaped emission hole along the
western portion of the remnant.  Furthermore, there appears to be {\it IRAS}
emission associated with other X-ray faint regions of G156, including its
southern limb, a region near the X-ray/optical bulge in the NE, and a NE--SW
oriented stripe near the NW edge of the remnant.  However, in a few spots, some
60~$\mu$m emission also appears to be associated with optical emission from
G156.  In particular, a faint ring of emission (difficult to see in
Figure~\ref{xray_iras}) appears to correlate with the H$\alpha$ Balmer
filaments, and the NE X-ray emission bulge can also be seen as an emission
feature in the 60~$\mu$m image.  

Such confusion between interstellar dust and SNR emission is not present in
Figure~\ref{xray_Av}, which shows the {\it ROSAT} X-ray image (blue) the
H$\alpha$ mosaic (green) and the 6' resolution $A_V$ extinction map (red) of
\citet{Dobashi05}.  Since this $A_V$ map is based solely on a star count
analysis of data from the Digitized Sky Survey, there can be no confusion with
line emission features.  However, once again the interstellar cloud near the NE
filament complex and the large triangular region in the SW are seen coincident
with high extinction zones in the $A_V$ map, as are the absorbing regions along
the remnant's southern limb and the NE--SW stripe along the northern limb.

The observed coincidences between X-ray emission holes and dusty interstellar
clouds strongly indicate that the clouds lie in front of the remnant creating
X-ray absorption shadows on the resulting X-ray image of the G156 remnant.  In
fact, many of the structures seen in the {\it ROSAT} image can be attributed to
absorption by these foreground clouds.  In addition, several radiative optical
emission features appear close to these obscuring dust regions, suggesting that
G156 may be physically interacting with these clouds leading to the relatively
bright radiative shock emission observed.  It should be noted, however, that
the arc of patchy emission stretching from near the NE filament complex down
toward the SW ridge of H$\alpha$ emission does not appear to be associated with
dust features in the 6510~\AA\ optical continuum image, in the 60~$\mu$m {\it
IRAS} image, or in the $A_V$ extinction map, although it seemingly does run
parallel to a stripe of fainter emission in the {\it ROSAT} X-ray image.

\section{Discussion}

Our mosaic H$\alpha$ image of G156 (Fig.\ 1) shows a rather surprising amount
of optical shock emission for a remnant previously reported to exhibit no
detectable optical emission.  G156's detected optical emission structure now
ranks it among the most extensive seen among galactic remnants, albeit at the
faint end of surface brightness.  The lack of visibility of G156 on the Palomar
Sky Survey is not particularly surprising given the broad passband of its
images. Nor is it surprising that the SNR emission was not detected by
\citet{Stecklum04} who imaged a region around the Bok globule CB 26 close to
the northwest filament complex in both H$\alpha$ and [S~II], but were limited
by a small ($2'$) field of view.  But the lack of detection by the narrow
passband filter galactic plane survey of \citet{Parker79} is puzzling. That
survey's l=155.5 $\degr$, b=+5 $\degr$ plates are well centred on G156 and yet
shows no discernible H$\alpha$, [S~II] $\lambda\lambda$6716,6731, or [O~III]
$\lambda$5007 emission.

G156's detected optical emission consists of a considerable amount of both
radiative and non-radiative filaments, something that is not commonly seen in
SNRs.  Only a handful of galactic remnants show such extensive non-radiative,
Balmer-dominated emission filaments alongside considerable radiative
filamentary nebulae. The few comparable
objects include the Cygnus Loop, with its bright radiative west and east nebulae
(NGC 6960 \& 6995) and extensive Balmer-dominated filaments outlining much of the
remnant's limb \citep{Levenson98}, and RCW 86 with its lone but bright
radiative emission SW limb region and non-radiative filaments outlining
most of the rest of the remnant \citep{Smith97}.

The mixture of both radiative and non-radiative optical emission suggests that
the remnant is situated in a highly non-homogenous ISM, with the radiative
emission associated with relatively dense cloud--SNR interactions while the
thin non-radiative filaments mark regions of much lower interstellar density.
However, \citet{Pfeff91} reported no spectral variations in four equally spaced
sectors of the remnant, concluding that there was no significant variation in
interstellar absorption across G156.  They did find a range of N(H) column
density of $9 \times 10^{20}$ to $ 2 \times 10^{22}$ cm$^{-2}$ depending on the
choice of model assumed, and chose the lower value in their final analysis.
\citet{Yama00} estimated N(H) = $3-4 \times 10^{21}$ cm$^{-2}$ using NEI plus
thermal or power law models.

Neither study discussed the possibility of N(H) variations, and yet the
presence of these large interstellar clouds along the line of sight clearly
indicates that the foreground column density varies significantly across the
face of the remnant. This complicates the assessment of the X-ray results, as
the foreground density is position dependent on scales smaller than the
effective apertures used in these studies.  A re-analysis of the {\it ROSAT}
data is certainly warranted, using emission primarily from the relatively
unobscured east side of the remnant.

The projected proximity and apparent alignment of the remnant's brighter
radiative filaments with several foreground interstellar clouds and dust lanes
hint at a possible physically interaction between the G156.2+5.7 remnant and
these interstellar clouds. The estimated distance to the T-Tauri star V347 Aur
associated with one of the southwestern foreground clouds is 250 -- 300 pc
\citep{Cohen78}, while the estimated distance to the Bok globule CB $26$ near
the remnant's bright northwest filament complex ranges from 140 to 300 pc
\citep{LH97,LS01}.  

If G156 is indeed interacting with part of the Taurus--Auriga cloud complex in
this direction, then the G156 remnant may be only $\approx$ 0.3 kpc distant,
significantly closer than the previous X-ray derived estimates of 1.3 -- 3 kpc.
It would also be much younger than the 15,000 -- 26,000 yr ages earlier
analyses have suggested.  In fact, assuming G156 is in the Sedov-Taylor
(adiabatic) expansion phase of its evolution, lies at a distance of just 0.3
kpc, is expanding in an intercloud density of 0.2 cm$^{-3}$
\citep{Yama99,Yama00}, and was generated by a SN with energy around the
canonical value of of $1 \times 10^{51}$ erg, then it has a radius of $\approx$
5 pc, an age of $\approx$ 400 yr, and a shock velocity around 5000 km s$^{-1}$.
Such a young age seems highly unlikely, as it should then have been seen as a
historical supernova. 

This extreme age problem is reduced somewhat if either the distance to the
clouds is somewhat farther, or the ambient ISM density is higher.  For example,
at a distance of 0.6 kpc, a Sedov analysis of G156 would give an age of
$\approx$ 2200 yr, and a shock velocity around 1500 km s$^{-1}$.  Such an age
would make G156 more akin to the SN~1006 SNR than the Cygnus Loop.  Its nearly
filled X-ray morphology is certainly unlike that seen in moderately old SNRs
and an age of only a few thousand years would be more consistent with the
observed presence of extensive non-radiative shock filaments which are most
often seen associated with high-velocity shocks in young remnants. A relatively
youthful remnant might also help explain the apparent enhanced abundances of Si
and S seen in the X-ray spectra of the centre of G156 \citep{Yama99}.  

On the other hand, the close apparent placement of the foreground clouds and
the radiative emission could be simply a coincidence.  Obviously, further
investigations into the distance and general properties of this SNR are needed.
For example, if its shock velocity is $\sim$ 1000 -- 2000 km s$^{-1}$, then
high resolution spectra of its Balmer dominated filaments should show evidence
for a broad H$\alpha$ emission component like that seen in the filaments in
Tycho and SN~1006.  Moreover, these Balmer dominated filaments should exhibit
large proper motions of order 1 arsec yr$^{-1}$. Even null results would
provide useful lower limits to the distance to G156, and hence lower limits on
the age of the SNR.   Whatever the outcome of future studies, G156 is clearly
an interesting remnant, with a now recognized extensive and diverse optical
emission structure.


\section*{Acknowledgments}
We thank J. Thorstensen for help with the optical coordinate matching.
CLG is supported through UK PPARC grant PPA/G/S/2003/00040.

\bsp
\label{lastpage}
\end{document}